\documentclass[12pt]{article}

\usepackage{graphics,cite,amssymb,epsfig,float,psfrag}
\usepackage[usenames,dvips]{color}
\usepackage{rotating}

\newcommand{\lsim}{\raisebox{-0.13cm}{~\shortstack{$<$ \\[-0.07cm] $\sim$}}~} 
\newcommand{\gsim}{\raisebox{-0.13cm}{~\shortstack{$>$ \\[-0.07cm] $\sim$}}~} 
 
\newcommand{\tb}{\tan\beta} 
\newcommand{\beq}{\begin{eqnarray}} 
\newcommand{\eeq}{\end{eqnarray}}

\begin{document}

\vspace{1cm}

\hfill LPT--ORSAY--10/96

\hfill CERN--PH--TH/2010--296

\vspace*{1.5cm}

\begin{center}

{\large\bf Revisiting the constraints on the Supersymmetric Higgs sector}

\vspace*{.1cm}

{\large\bf at the Tevatron}

\vspace{.8cm}

{\large Julien Baglio$^{1}$ and Abdelhak Djouadi$^{1,2}$} 

\vspace*{8mm}

$^1$ Laboratoire de Physique Th\'eorique, Universit\'e Paris XI et CNRS,
F-91405 Orsay, France.\\
$^2$ Theory Unit, CERN, 1211  Gen\`eve 23, Switzerland.
\end{center}

\vspace{1.4cm}

\begin{abstract} 
We analyze the production of the neutral Higgs particles of the Minimal
Supersymmetric extension of the Standard Model at  the Fermilab Tevatron
collider.  We consider the two main production and detection channels: 
gluon--gluon and bottom quark fusion leading to Higgs bosons which subsequently
decay into tau leptons, $gg, b\bar b\! \to\! {\rm Higgs}\! \to\!
\tau^+\tau^-$. We update the production cross sections  and the decay branching
ratios and obtain production rates that are significantly smaller at high
masses than the ones used by the CDF and D0 experiments in their search. We
then evaluate the various theoretical uncertainties that affect these rates,
uncertainties that have not been been considered in the CDF/D0  analyses
and which turn out to be rather large. Including these two effects will
significantly loosen the constraints obtained on the supersymmetric Higgs
sector at the Tevatron.

\end{abstract} 

\thispagestyle{empty}

\newpage
\setcounter{page}{1}

The search for the Higgs bosons, the remnants of the spontaneous breaking of the
electroweak symmetry that is at the origin of the elementary particle  masses,
is the main goal of present high--energy colliders. While a single Higgs boson
is predicted in the Standard Model (SM), the minimal realization of the symmetry
breaking with only one Higgs doublet field \cite{Higgs}, the  Higgs sector is
extended in supersymmetric theories \cite{SUSY}, that are widely considered to
be the most attractive extensions of the SM as they stabilize the hierarchy
between the electroweak and Planck scales induced by the large radiative
corrections to the Higgs boson mass. In the minimal extension, the Minimal
Supersymmetric Standard Model (MSSM) \cite{SUSY}, two Higgs doublet fields are
required, leading to the existence of five Higgs particles: two CP--even $h$ and
$H$, a CP--odd $A$ and two charged $H^\pm$ particles \cite{HHG,Review}.

In the MSSM, two parameters are needed to describe the Higgs sector at
tree--level: the mass $M_A$ of the pseudoscalar boson and the ratio of vacuum
expectation values of the two Higgs fields, $\tan\beta$, that is expected to
lie in the range $1 \lsim \tb \lsim 50$. At high $\tb$ values, $\tb \gsim 10$,
one of the neutral CP--even states has almost exactly the properties of the SM
Higgs particle: its couplings to fermions and gauge bosons are the same, but
its mass is restricted to values $M_h^{\rm max}  \approx 110$--135 GeV
depending on some SUSY parameters that enter the radiative corrections
\cite{Review}. The other CP--even and the CP--odd states, that we will denote
collectively by $\Phi\!=\!A,H(h)$, are then almost degenerate in mass and have
the same properties: no couplings to gauge bosons, while the couplings to
isospin down--type (up--type) quarks and charged leptons are (inversely)
proportional to $\tb$.\vspace*{-.5mm}

Thus, for $\tb\! \gsim\! 10$, the $\Phi$ boson couplings to bottom quarks and
$\tau$--leptons are strongly enhanced while those to top quarks are
suppressed.  As a result, the phenomenology of these states becomes rather
simple. To a very good approximation, the $\Phi$ bosons decay almost
exclusively into $b\bar b$ and $\tau^+\tau^-$ pairs with branching ratios of,
respectively, $\approx\! 90\%$ and $ \approx\! 10\%$, while the other decay
channels are suppressed to a negligible level \cite{HDECAY}.  The main
production mechanisms for these particles are those processes which involve the
couplings to bottom quarks. At hadron colliders,  these are the gluon--gluon
fusion mechanism, $gg \to \Phi$, which dominantly  proceeds through $b$--quark
triangular loops \cite{ggH-LO,SDGZ}  and  bottom--quark fusion, $b\bar b \to
\Phi$  \cite{bbH-LO,bbH-NLO,bbH-NNLO}, in which the bottom quarks are directly
taken from the protons in a five active flavor scheme \cite{F-bbA}. The  latter
process is similar to the channel $p\bar p \to b\bar b\Phi$ when no $b$--quarks
are detected in the final state \cite{F-bbA}.

With its successful operation in the last years, the Fermilab Tevatron 
collider has now collected a substantial amount of data which allows the CDF
and D0 experiments to be sensitive to the MSSM Higgs sector.  Stringent
constraints beyond the well established LEP bounds $M_{A}, M_h \gsim M_Z$ and
$\tb \gsim 3$ \cite{LEP-Higgs}, have been set on the MSSM parameter space
$[M_A, \tb$] using the process $gg, b\bar b \to \Phi \to \tau^+ \tau^-$.
Moderate $A$ masses,  $M_A\! \approx\! 100$--200 GeV, together with high $\tb$
values, $\tb\! \gsim\! 30$, have been excluded at the 95\% confidence level
(CL) \cite{Tevatron-MSSM0,Tevatron-MSSM}.

Nevertheless, a very important issue has been overlooked in these experimental 
analyses: the theoretical uncertainties that affect the production and decay
rates, which can be important despite of the fact that some higher order 
perturbative corrections to these processes are known. These are mainly due to
the unknown higher order corrections in perturbation theory, the still not
satisfactory parametrization of the parton distribution functions (PDFs), as
well as the parametric uncertainties stemming from the not very precisely 
measured values of the strong coupling constant $\alpha_s$ and the bottom quark
mass $M_b$. In a recent analysis \cite{Hpaper}, it has been shown that these
uncertainties can be rather large  at the early stage of the CERN large Hadron
Collider (lHC), in much the same way as in the case of SM Higgs production both
at the lHC \cite{Hpaper} and at the Tevatron \cite{Hpaper0}.

In this Letter, we first update the decay branching ratios and production cross
sections of the $\Phi$ bosons and find that the latter are significantly lower
at high masses than the ones assumed in the Tevatron analyses.  We then evaluate
the theoretical uncertainties that affect these rates and find them to be very
large, possibly lowering the cross sections times branching ratios by a factor
$\approx 2$. When included in the D0 and CDF combined analysis of the MSSM Higgs
bosons \cite{Tevatron-MSSM}, the correct normalization and the  uncertainties
will drastically reduce the MSSM  $[M_A, \tb]$ parameter space that has been
excluded.


For the evaluation of the cross sections in the $gg \to \Phi$ and $b\bar b  \to
\Phi$ production processes at the Tevatron, we will concentrate on the
pseudoscalar $A$ case and follow very closely the recent analysis performed in
Ref.~\cite{Hpaper} for the lHC. We calculate $\sigma (gg \to A)$, known up
next-to-leading order (NLO) only \cite{SDGZ},  using the program {\tt HIGLU}
\cite{Michael} with central values for the renormalization and factorization
scales, $\mu_R\!=\!\mu_F\!=\!\mu_0\!= \!\frac12 M_A$; only the dominant loop
contribution of the bottom quark is taken into account.  For the  $b\bar b\to
A$ process, known  up to next-to-next-to-leading order (NNLO) \cite{bbH-NNLO},
we use the program {\tt bbh@nnlo}\footnote{We thank R. Harlander for
  providing us with his code.}  with a central scale
$\mu_R\!=\mu_F\!= \!\mu_0\!=\! \frac14 M_A$. In both cases, we work in the
$\overline{\rm MS}$ scheme for the renormalization  of the $b$--quark mass;
however, while $\overline{m}_b (\overline{m}_b)$ is  used in the $gg$ process, 
$\overline{m}_b(\mu_R)$ is adopted in the $b\bar b$ channel. The resulting
partonic cross sections are then folded with the latest MSTW sets of PDFs
\cite{MSTW}, consistently at the respective orders, NLO or NNLO, in
perturbation theory.

In both the $gg \to A$ and $b\bar b  \to A$ processes, we assume the $b\bar bA$
coupling to be SM--like, $\lambda_{Abb}= m_b/v$. To obtain the true cross 
sections, one has to rescale the numbers which will be given by a factor 
$\tan^2 \beta$. In addition, to obtain the cross section for both the $A$ and
$H(h)$ bosons, an  additional factor of two has to be included. As a
consequence of chiral symmetry for $M_\Phi\! \gg\!\overline{m}_b$ and since 
the $H(h)$ masses and couplings are very close to those of $A$, this represents
an excellent approximation\footnote{Note that there are additional
  SUSY contributions in $gg\!\to\! H/h$ which do not appear in
  $gg\!\to\! A$ but they are very small for a
large SUSY breaking scale, $M_\Phi \ll M_S$ \cite{SUSY-QCD-pap}. Furthermore,
there are one--loop  vertex corrections to the $\Phi b\bar b$ coupling  due to
SUSY particles which can be significant as they grow with $\tan\beta$
\cite{Deltab}. They are implemented in the major codes which calculate the
MSSM Higgs spectra and can be readily included. However, in the case of $p\bar
p\! \to\! \Phi\! \to\! \tau^+ \tau^-$, they almost entirely cancel in the
cross section times branching ratios and the remaining part is so small that
it has no practical impact whatever benchmark scenario is considered. This can
be seen from the almost identical tables XI--XIV and Figs.~4 of
Ref.~\cite{Tevatron-MSSM} that describe four benchmark scenarios
\cite{benchmark}.}.

The results for the cross sections $\sigma(gg\!\to\! A)$ and $\sigma(b\bar b\!
\to\!  A)$ are shown in the main frames of Fig.~1 for the Higgs mass range that
is relevant at the Tevatron, $M_A=90$--200 GeV. We have compared our values
with those given by the program that has been used by the CDF and D0
collaborations for their cross section normalization, {\tt FeynHiggs}
\cite{FeynHiggs}. This program, initially supposed to only provide precise
values for the MSSM Higgs masses and couplings, gives also grids for production
cross sections which should be used with care. For the $b \bar b\!\to\! A$
channel, we obtain cross sections that are $\approx 30\%$ smaller. The reason
is that  {\tt FeynHiggs} simply provides the values given in the original paper
\cite{bbH-NNLO} which uses the outdated MRST2002 set of PDFs which are only
partly at NNLO. In the case of $gg\to A$, the agreement is better as we obtain
a cross section that is only $\approx 10\%$ higher; this can be attributed to
the different central scale and renormalization scheme for $M_b$
that have been used\footnote{We thank S. Heinemeyer for a discussion on these points.}.

\begin{figure}[!h] 
\begin{center} 
\mbox{
\epsfig{file=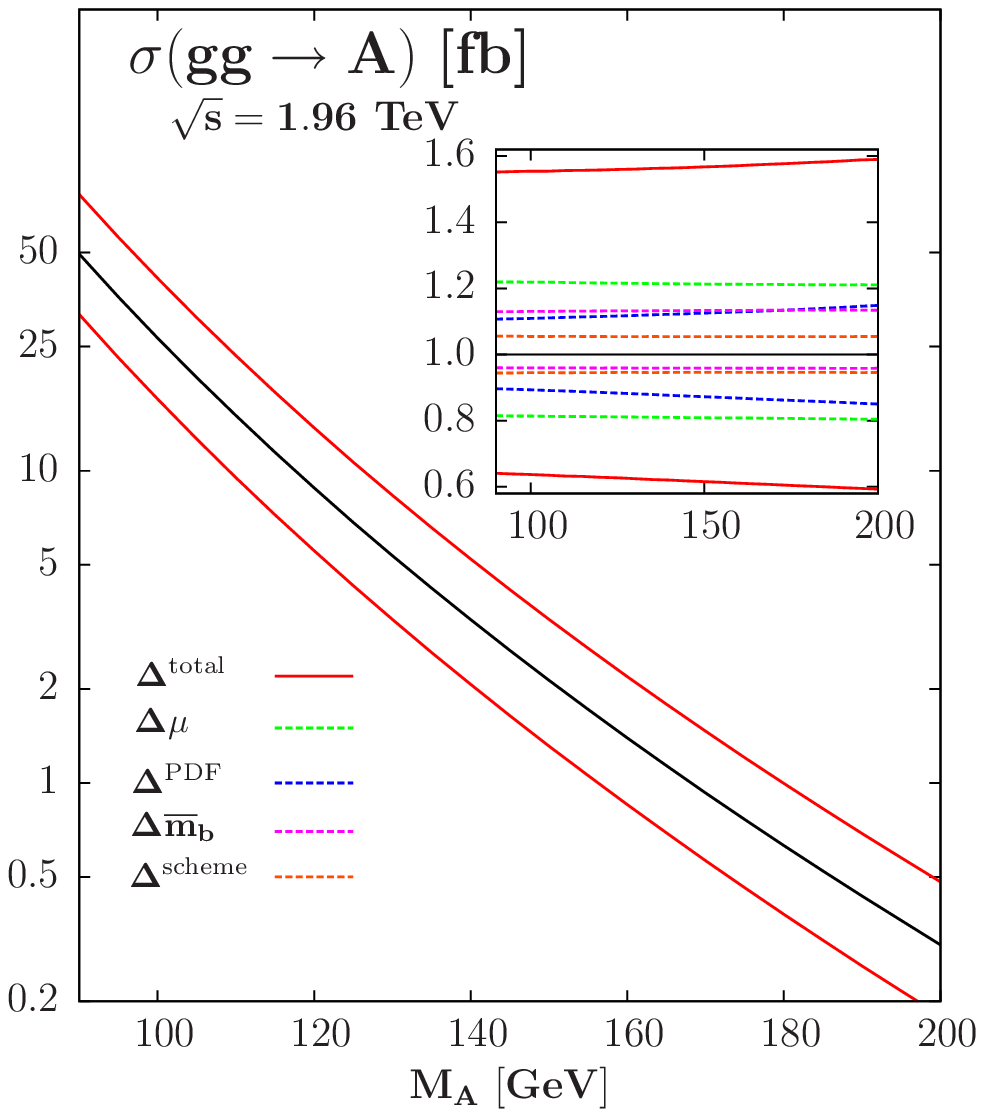,width=8.2cm,height=9cm}\hspace*{1mm}
\epsfig{file=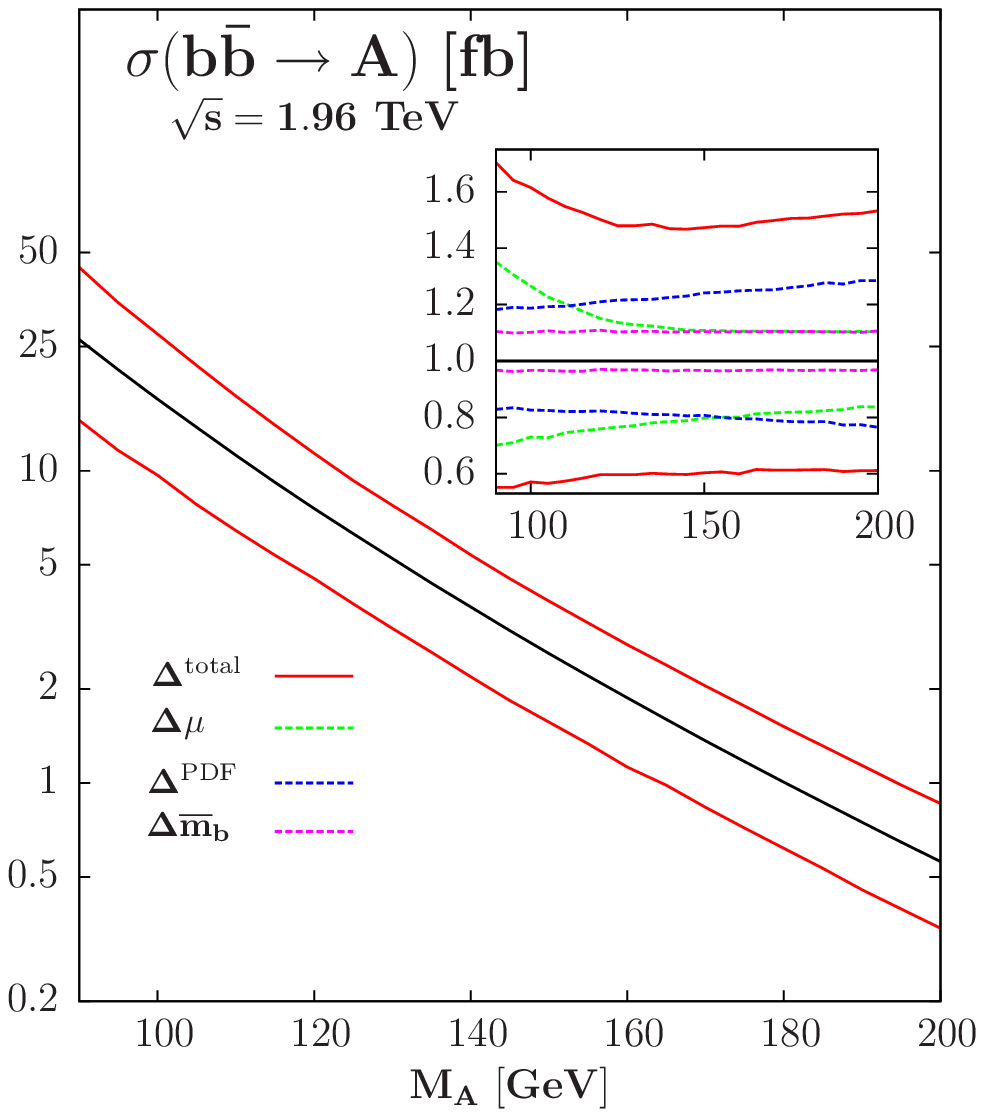,width=8.2cm,height=9cm}  } 
\end{center} 
\vspace*{-5mm}
\caption[]{The normalization of the cross sections $\sigma^{\rm NLO}_{gg\!\to
\!A}$ (left) and $\sigma^{\rm NNLO}_{b\bar b \!\to\!A}$ (right) at Tevatron 
energies as a function of $M_A$ when using the MSTW PDFs and unit $Ab\bar b$ 
couplings. In the inserts, shown are the various sources of theoretical 
uncertainties when the rates are normalized to  the central values.}
\vspace*{-3mm}
\label{Errors}
\end{figure}

For the evaluation of the theoretical uncertainties that affect the Higgs 
production cross sections as well as the decay branching ratios,  we will
proceed as follows. 

The Higgs decays branching ratios (BR) have been discussed in
Ref.~\cite{Hpaper} and are simply affected by the parametric uncertainties on
the input $b$--quark mass and $\alpha_s$. As the QCD corrections to the
dominant $\Phi  \to b\bar b$ decays are large, they are resummed by switching
from the $b$--quark pole mass $M_b$ which appears at tree--level to the running
quark mass in the $\overline{\rm MS}$   scheme evaluated at the scale of the
Higgs mass, $\overline{m}_b(M_\Phi )$. The uncertainties come then from: $i)$
the starting  point for the $b$--quark mass $\overline{m}_b(\overline{m}_b)=
4.19^{+0.18} _{-0.06}$ GeV \cite{PDG} where the central value corresponds to a
pole mass of $M_b=4.71$ GeV; and $ii)$ the error on the QCD coupling
$\alpha_s(M_Z^2)\!=\! 0.1171 \pm 0.0014$ at NNLO (the value adopted in the
cross sections) \cite{MSTW}  which is used to run the $b$--quark mass from
$\overline{m}_b$ up to $M_\Phi$. Assuming that there is no uncertainty in the
$\Phi  \to \tau^+ \tau^-$ decay as only electroweak effects are present, and 
adding the errors in quadrature, one finds an uncertainty of $\approx
+4\%,-9\%$ on BR$(\Phi\!  \to\! \tau^+\tau^-)$ and  $\approx +1\%,-0.5\%$ on
BR$(\Phi  \! \to \! b\bar b )$ at the 1$\sigma$ level over the entire  relevant
Higgs mass range, $M_A=90$--200 GeV.

In the case of the production cross sections, the uncertainty from the missing
higher orders in perturbation theory is usually estimated by varying the
renormalization and factorization scales in the domains $ \mu_0/\kappa \le
\mu_R,\mu_F  \le \kappa \mu_0$ around the central scales $\mu_0$, with the
additional restriction $1/\kappa \le \mu_R/\mu_F \le \kappa$ imposed. 
While we choose $\kappa\!=\!2$ for the $gg\to A$ process,  the value
$\kappa\!=\!3$ is adopted for $b\bar b \to A$. The reason is that it is well
known that the cross sections in $b\bar b \to A$ and in the twin process
$q\bar q, gg \to b\bar b A$ in a four--flavor scheme differ significantly 
\cite{bbH-diff} and by allowing for a wider domain for scale variation  and,
hence, a larger scale uncertainty, the two results become more consistent with
each other. Furthermore, in the $gg\to A$ process, there is an additional
uncertainty that we will consider: the one due to the choice of the scheme for
the renormalization of the $b$--quark mass. The latter is estimated by taking
the difference between the results obtained in the on--shell mass and
$\overline{\rm MS}$ schemes and allowing for both signs. The inclusion of this
additional effect is similar to increasing the  domain of scale variation from
$\kappa\!=\!2$ to $\kappa\!=\!3$.

For the combined uncertainties from the PDFs and the QCD coupling $\alpha_s$, we
will use the scheme made available by the MSTW collaboration \cite{MSTW}. The
PDF+$\Delta^{\rm exp}\alpha_s$  uncertainty, with  $\alpha_s (M_Z^2)\!=\! 0.120
\pm 0.002$ at NLO for $gg \to A$ and   $\alpha_s(M_Z^2)\!=\!0.1171 \!\pm\!
0.0014$ at NNLO for $b\bar b \to A$, is evaluated at the 90\% CL. To this, we
add in quadrature the impact of a theoretical error on $\alpha_s$, estimated by
MSTW to be $\Delta^{\rm th} \alpha_s\! \approx\! 0.003$ at NLO and $\Delta^{\rm
th}\! \approx\! 0.002$ at NNLO.  Finally, in the case of the $b\bar b \to A$
process, there is an effect induced by the uncertainty in the value of $M_b$  in
the $b$--quark density. This effect is again estimated within the MSTW scheme by
allowing  for an uncertainty on the pole $b$--mass of $\pm 0.25$ GeV from
the MSTW central value $M_b=4.75 \pm 0.25$ GeV. This resulting uncertainty is
also added in quadrature to the PDF+$\Delta^{\rm exp}\alpha_s$ one.  Note that
we have evaluated both cross sections with four other PDF sets and found that
the maximal  values are obtained with MSTW in $gg\!\to\! A$, while some other
schemes give $\approx\  20\%$ lower rates.

Finally, there is the uncertainty on the $b$--quark mass that affects the 
amplitudes of these processes, in much the same way as what has been  
discussed for the Higgs decay branching ratios. It is estimated by evaluating
the maximal values of the cross sections when one includes the error on the
input $\overline{\rm MS}$ $b$--quark mass at the scale $\overline{m}_b$,
$\overline{m}_b (\overline{m}_b)=4.19^{+0.18}_{-0.06}$  GeV, and in the case
of the $b\bar b \to A$ process where the Yukawa coupling is defined at the
high scale, $\propto \overline{m}_b (\mu_R)$,  the error on the coupling 
$\alpha_s(M_Z^2)\!=\!0.1171\! \pm\! 0.0014$ at NNLO (in this case, it is almost
the same uncertainty as in the $A \to b\bar b$ decay).

The results for these uncertainties on the production cross sections at the
Tevatron are shown in Fig.~\ref{Errors} for the $gg\!\to\! A$ and $b\bar b 
\!\to \!A$ processes as a function of $M_A$. In the $gg$ case and almost
independently of $M_A$, the scale variation in a domain with $\kappa=2$  leads
to an uncertainty  ${\cal O}(\pm 20\%)$, while the  uncertainty from the scheme
dependence in the renormalization of $M_b$ is about $\pm 6\%$; they add up to
$\approx 25\%$ that is only slightly lower than the scale uncertainty in  the
$b\bar b$  process, $\approx 30\%$ for low $M_A$,  in which the domain of
variation is extended to $\kappa=3$. In the $gg\to A$ ($b\bar b \to A)$
channel, the PDF+$\Delta^{\rm exp+th} \alpha_s$ (with $\Delta M_b$ in addition
for $b\bar b \to A$) uncertainties are at the level of $\pm 10\%$ ($\pm 20\%$)
for $M_A \approx 100$ GeV  and larger ($\pm 30\%$ in $b\bar b\!\to\! A$) at
$M_A \approx 200$ GeV where the more uncertain high Bjorken--$x$ values for the
gluon and bottom quark densities are probed. The parametric error on
$\overline{m}_b$ leads to a $\approx +13\%, -4\%$ uncertainty in the $gg\to A$
process and slightly less in the case of $b\bar b \to A$.

We turn now to the issue of combining these uncertainties. Clearly, the scale and
scheme uncertainties, which are purely theoretical and both emerge from the truncation
of the perturbative series, should be added linearly. The PDF+$\Delta \alpha_s$+$\Delta
m_b$ uncertainty, that we also would like to view as a reflection of the theoretical
ambiguities due to the parametrization  of the PDFs and which have no statistical
ground, will be evaluated on the minimal and maximal values of the cross sections with
respect to scale and scheme variation (see Ref.~\cite{Hpaper} for the argumentation).
This procedure gives results that are similar to those obtained with a linear addition
of the scale+scheme and PDF+$\Delta \alpha_s$+$\Delta m_b$ uncertainties as advocated 
in, for instance, Ref.~\cite{LHCXS}. 
Finally, at this stage, we add linearly the parametric uncertainty on $\overline{m}_b$
which, in the case of interest, will drop anyway in the final result (see below). 

The combined uncertainties on the cross sections, when using this procedure,
are also shown in Fig.~\ref{Errors} for the two production channels. As can be
seen, they are very large: at low Higgs masses, $M_A\approx 100$ GeV, one has 
$\approx +55\%,-35\%$ for  $\sigma( gg \to A)$   and $\approx +60\%,-40\%$ for
$\sigma(b\bar b \to A)$ which become at  masses $M_A\approx 200$ GeV,
respectively, $\approx +60\%,-40\%$ and $\approx +50\%,-40\%$.

\begin{figure}[!h]
\begin{center}
\vspace*{2mm}
\epsfig{file=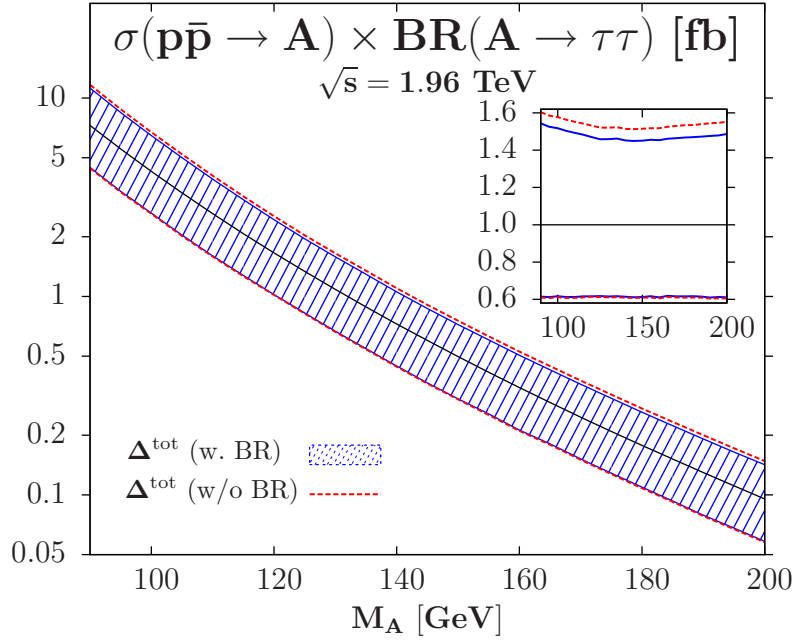,scale=0.99} 
\end{center}
\vspace*{-4mm}
\caption[]{$\sigma(p\bar p\! \to\! A)\times 
{\rm BR}(A \!\to\! \tau^+ \tau^-)$ as a function of $M_A$ at the Tevatron, 
together with the associated overall theoretical uncertainty;  the uncertainty 
when excluding that on the branching ratio is also displayed.  In the inserts,
shown are the  relative deviations from the central values.}
\vspace*{-1mm}
\label{Total}
\end{figure}

To consider the final state topology that has been searched for by the  D0 and
CDF collaborations \cite{Tevatron-MSSM}, i.e.  $p\bar p\! \to\! {\rm Higgs}\!
\to\! \tau^+\tau^-$,  one has first to add the cross sections for the two
channels $gg \to A$ and $b\bar b \to A$, and then to multiply the resulting
production cross section by the Higgs branching ratio BR($A \to \tau^+\tau^-)
\approx 10\%$. The resulting $\sigma(p\bar p\!\to \! A)\!\times\! {\rm
BR}(A\!\to\! \tau^+ \tau^-)$ at the Tevatron is shown in  Fig.~\ref{Total} as a
function of $M_A$. We stress again that to obtain the true rate for
$\Phi\!=\!A\!+\!H(h)$, one  has to multiply the given values by a
factor $2\tan^2\beta$.

In Fig.~\ref{Total}, shown also are the associated overall theoretical
uncertainties.  The uncertainty from the cross section alone is dominated by
that of  $gg\!\to\!A$ at low Higgs masses and $b\bar b\! \to\! A$ at high
masses as the corresponding cross sections are largest. In the product
$\sigma(p\bar p \to A) \times {\rm BR}(A\to \tau^+ \tau^-)$, the parametric
uncertainty that  its common to the production and decay rates almost cancels
out as shown by the solid curves in Fig.~\ref{Total} and only a few percent are
left. This leads to a smaller uncertainty in $\sigma (p\bar p \!\to \! A)
\times {\rm BR}(A\!\to\! \tau^+\tau^-) $ than in $\sigma (p\bar p \!\to \!A)$
alone. The final theoretical uncertainty for $p\bar p \!
\to\! A \! \to\! \tau^+ \tau^-$ at the Tevatron is of order  $+50\%,
-40\%$.

\begin{figure}[!h]
\begin{center}
\vspace*{.2mm}
\epsfig{file=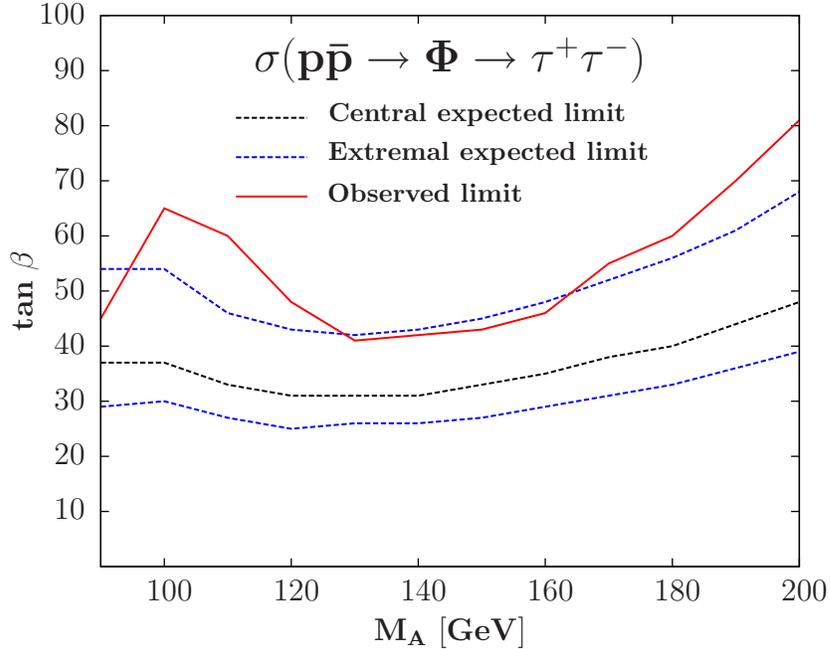,scale=0.99} 
\end{center}
\vspace*{-5mm}
\caption[]{Contours for the expected $\sigma(p\bar p\! \to\! \Phi \!\to 
\!\tau^+ \tau^-)$ rate at the Tevatron in the [$M_A, \tb$] plane with the 
associated theory  uncertainties, confronted to the 95\% CL exclusion limit.} 
\vspace*{-.2mm}
\label{Scan}
\end{figure}

To illustrate the impact of these theoretical uncertainties on the MSSM [$M_A,
\tan\beta$] parameter space that is probed when searching experimentally for
the  $p\bar p \! \to \! \Phi \! \to \! \tau^+ \tau^-$ channel, we show in
Fig.~\ref{Scan}  the contour of the cross section times branching ratio in this
plane, together with the contours when the uncertainties are included. We  apply
the model independent 95\%CL expected and observed limits from the CDF/D0
analysis  (Table X of Ref.~\cite{Tevatron-MSSM}). However, rather than applying
the limits on the central $\sigma \times$BR rate, we  apply them on the minimal
one  when the theory uncertainty is included. Indeed, since the latter  has a
flat prior,  the minimal $\sigma \times$BR value is as respectable and likely as
the central value.  One observes then that only values $\tb\gsim 50$ are 
excluded in the mass ranges, $M_\Phi \!\approx \!95$--125 GeV and $M_\Phi\!
\gsim \!165$ GeV. In the intermediate range $M_\Phi \! \approx \! 125$--165 GeV, the
exclusion  limit is $\tb\! \gsim\! 40$--45, to be contrasted with the values $\tan
\beta\! \gsim\! 30$ excluded in the CDF/D0 analysis. Hence, the inclusion of the
theory uncertainties has a drastic impact on the allowed  [$M_A, \tan\beta$]
parameter space. 

Finally, let us note that there is a subleading channel which has also been 
considered at the Tevatron, $bg\! \to\! \Phi b\! \to\! 3b$ \cite{Tevatron-3b}.
The evaluation of the theory uncertainties in $bg\! \to\! \Phi b$  
\cite{bbH-NLO} follows that of the parent process $b\bar b\! \to\! \Phi$ (for
which it is part of the NLO contributions) and similar results, i.e. a total
uncertainty of $\approx \pm 40\%$, are expected. However, in this case, it is
the $\Phi\to b\bar b$ decay which is considered experimentally and, since
BR$(\Phi \! \to \! b\bar b)$ has a small error, the uncertainties in $\sigma $
and $\sigma \times {\rm BR}$ are almost the same. These uncertainties will thus
also impact the excluded  [$M_A, \tb$]  parameter space\footnote{Note
  that the $2\sigma$ excess observed by CDF in this channel
cannot be a Higgs signal as it would correspond to a much larger excess in
$p\bar p \to \tau^+\tau^-$ which has not been observed. In addition,
this $3b$ channel probes huge $\tb\! \gsim\! 100$
values, whereas values $\tb\!\gsim\!50$ are theoretically not favored as they
lead to a $\Phi b\bar b$ Yukawa coupling that is non perturbative. For such
values, the total Higgs widths are to be included.}.

In conclusion, we have updated the cross sections for the production of  the
MSSM CP--odd like Higgs bosons $\Phi$ at the Tevatron in the processes $gg\to
\Phi$ and $b\bar b \to \Phi$ and found smaller rates in the high Higgs mass 
range compared to those assumed by the CDF and D0 experiments. We have then
evaluated the associated theoretical uncertainties,  including also the ones in
the $\Phi\!\to \!\tau^+ \tau^-$ decay branching fractions, and find that they
are very large. These uncertainties, together with the correct normalization,
affect significantly the exclusion limits set on the MSSM parameter space from
the negative Higgs searches in the channel $p\bar p \to \Phi \to \tau^+ \tau^-$
at the Tevatron.  Additional material is given in Tables 1 and 2 which summarize
our results.\bigskip

{\bf Acknowledgments:} Discussions with Michael Spira are gratefully 
acknowledged. This work is supported by the European network HEPTOOLS.

\newpage

\begin{table}[!h]{\small%
\let\lbr\{\def\{{\char'173}%
\let\rbr\}\def\}{\char'175}%
\renewcommand{\arraystretch}{1.3}
\vspace*{10mm}
\begin{center}
\begin{tabular}{|c||c|ccccc||cc|cc|}\hline
$M_A$ & $\sigma^{\rm NLO}_{\rm gg \to A}$ & scale & scheme &
PDFs & param & total & BR$(\tau\tau) \hspace*{-9mm}$ &  & $\sigma \times
$BR$ \hspace*{-9mm}$ &
 \\ \hline
$90$ & $49.46$ & $^{+22.0\%}_{-18.5\%}$ & $^{+5.6\%}_{-5.6\%}$ &
$^{+10.7\%}_{-10.3\%}$ & $^{+13.0\%}_{-4.0\%}$ & $^{+55.2\%}_{-35.9\%}$ &
$9.60$ & $^{+3.8\%}_{-8.8\%}$ & $4.75$ & $^{+49.7\%}_{-35.7\%}$ \\ \hline
$95$ & $36.07$ & $^{+22.0\%}_{-18.6\%}$ & $^{+5.6\%}_{-5.6\%}$ &
$^{+10.8\%}_{-10.5\%}$ & $^{+13.0\%}_{-4.0\%}$ & $^{+55.3\%}_{-36.1\%}$ &
$9.70$ & $^{+3.7\%}_{-8.8\%}$ & $3.50$ & $^{+49.9\%}_{-35.9\%}$ \\ \hline
$100$ & $26.62$ & $^{+21.9\%}_{-18.6\%}$ & $^{+5.5\%}_{-5.5\%}$ &
$^{+10.9\%}_{-10.7\%}$ & $^{+13.1\%}_{-4.0\%}$ & $^{+55.4\%}_{-36.3\%}$ &
$9.79$ & $^{+3.8\%}_{-8.8\%}$ & $2.61$ & $^{+50.0\%}_{-36.0\%}$ \\ \hline
$105$ & $19.90$ & $^{+21.9\%}_{-18.7\%}$ & $^{+5.5\%}_{-5.5\%}$ &
$^{+11.1\%}_{-10.8\%}$ & $^{+13.0\%}_{-4.0\%}$ & $^{+55.5\%}_{-36.5\%}$ &
$9.88$ & $^{+3.8\%}_{-8.8\%}$ & $1.97$ & $^{+50.1\%}_{-36.3\%}$ \\ \hline
$110$ & $15.03$ & $^{+21.8\%}_{-18.7\%}$ & $^{+5.5\%}_{-5.5\%}$ &
$^{+11.2\%}_{-11.0\%}$ & $^{+13.1\%}_{-4.1\%}$ & $^{+55.7\%}_{-36.7\%}$ &
$9.96$ & $^{+3.8\%}_{-8.8\%}$ & $1.50$ & $^{+50.3\%}_{-36.6\%}$ \\ \hline
$115$ & $11.46$ & $^{+21.7\%}_{-18.8\%}$ & $^{+5.5\%}_{-5.5\%}$ &
$^{+11.4\%}_{-11.2\%}$ & $^{+13.1\%}_{-4.0\%}$ & $^{+55.7\%}_{-36.9\%}$ &
$10.04$ & $^{+3.8\%}_{-8.8\%}$ & $1.15$ & $^{+50.4\%}_{-36.7\%}$ \\ \hline
$120$ & $8.82$ & $^{+21.6\%}_{-18.8\%}$ & $^{+5.5\%}_{-5.5\%}$ &
$^{+11.5\%}_{-11.4\%}$ & $^{+13.2\%}_{-4.1\%}$ & $^{+55.8\%}_{-37.1\%}$ &
$10.12$ & $^{+3.8\%}_{-8.8\%}$ & $0.89$ & $^{+50.5\%}_{-37.0\%}$ \\ \hline
$125$ & $6.84$ & $^{+21.6\%}_{-18.8\%}$ & $^{+5.4\%}_{-5.4\%}$ &
$^{+11.7\%}_{-11.7\%}$ & $^{+13.2\%}_{-4.1\%}$ & $^{+56.0\%}_{-37.3\%}$ &
$10.19$ & $^{+3.8\%}_{-8.8\%}$ & $0.70$ & $^{+50.6\%}_{-37.1\%}$ \\ \hline
$130$ & $5.35$ & $^{+21.5\%}_{-18.9\%}$ & $^{+5.4\%}_{-5.4\%}$ &
$^{+11.8\%}_{-11.9\%}$ & $^{+13.2\%}_{-4.1\%}$ & $^{+56.0\%}_{-37.6\%}$ &
$10.26$ & $^{+3.8\%}_{-8.7\%}$ & $0.55$ & $^{+50.9\%}_{-37.4\%}$ \\ \hline
$135$ & $4.21$ & $^{+21.5\%}_{-18.9\%}$ & $^{+5.4\%}_{-5.4\%}$ &
$^{+12.0\%}_{-12.1\%}$ & $^{+13.3\%}_{-4.1\%}$ & $^{+56.3\%}_{-37.8\%}$ &
$10.33$ & $^{+3.8\%}_{-8.8\%}$ & $0.43$ & $^{+51.0\%}_{-37.8\%}$ \\ \hline
$140$ & $3.34$ & $^{+21.4\%}_{-19.0\%}$ & $^{+5.4\%}_{-5.4\%}$ &
$^{+12.2\%}_{-12.3\%}$ & $^{+13.3\%}_{-4.1\%}$ & $^{+56.4\%}_{-38.0\%}$ &
$10.39$ & $^{+3.8\%}_{-8.7\%}$ & $0.35$ & $^{+51.2\%}_{-37.9\%}$ \\ \hline
$145$ & $2.66$ & $^{+21.4\%}_{-19.0\%}$ & $^{+5.4\%}_{-5.4\%}$ &
$^{+12.3\%}_{-12.5\%}$ & $^{+13.3\%}_{-4.1\%}$ & $^{+56.6\%}_{-38.2\%}$ &
$10.46$ & $^{+3.7\%}_{-8.8\%}$ & $0.28$ & $^{+51.3\%}_{-38.4\%}$ \\ \hline
$150$ & $2.13$ & $^{+21.3\%}_{-19.1\%}$ & $^{+5.4\%}_{-5.4\%}$ &
$^{+12.5\%}_{-12.8\%}$ & $^{+13.3\%}_{-4.1\%}$ & $^{+56.8\%}_{-38.4\%}$ &
$10.52$ & $^{+3.7\%}_{-8.8\%}$ & $0.22$ & $^{+51.5\%}_{-38.6\%}$ \\ \hline
$155$ & $1.72$ & $^{+21.3\%}_{-19.1\%}$ & $^{+5.4\%}_{-5.4\%}$ &
$^{+12.7\%}_{-13.0\%}$ & $^{+13.3\%}_{-4.1\%}$ & $^{+56.9\%}_{-38.7\%}$ &
$10.57$ & $^{+3.8\%}_{-8.7\%}$ & $0.18$ & $^{+51.8\%}_{-38.5\%}$ \\ \hline
$160$ & $1.40$ & $^{+21.2\%}_{-19.2\%}$ & $^{+5.4\%}_{-5.4\%}$ &
$^{+12.9\%}_{-13.2\%}$ & $^{+13.4\%}_{-4.1\%}$ & $^{+57.1\%}_{-38.9\%}$ &
$10.63$ & $^{+3.8\%}_{-8.7\%}$ & $0.15$ & $^{+52.0\%}_{-38.9\%}$ \\ \hline
$165$ & $1.14$ & $^{+21.2\%}_{-19.2\%}$ & $^{+5.4\%}_{-5.4\%}$ &
$^{+13.1\%}_{-13.4\%}$ & $^{+13.4\%}_{-4.1\%}$ & $^{+57.4\%}_{-39.1\%}$ &
$10.68$ & $^{+3.9\%}_{-8.7\%}$ & $0.12$ & $^{+52.3\%}_{-39.1\%}$ \\ \hline
$170$ & $0.93$ & $^{+21.2\%}_{-19.3\%}$ & $^{+5.4\%}_{-5.4\%}$ &
$^{+13.3\%}_{-13.6\%}$ & $^{+13.4\%}_{-4.1\%}$ & $^{+57.6\%}_{-39.3\%}$ &
$10.74$ & $^{+3.8\%}_{-8.8\%}$ & $0.10$ & $^{+52.4\%}_{-39.4\%}$ \\ \hline
$175$ & $0.76$ & $^{+21.2\%}_{-19.3\%}$ & $^{+5.4\%}_{-5.4\%}$ &
$^{+13.5\%}_{-13.9\%}$ & $^{+13.4\%}_{-4.2\%}$ & $^{+57.8\%}_{-39.6\%}$ &
$10.79$ & $^{+3.8\%}_{-8.7\%}$ & $0.08$ & $^{+52.7\%}_{-39.5\%}$ \\ \hline
$180$ & $0.63$ & $^{+21.1\%}_{-19.4\%}$ & $^{+5.4\%}_{-5.4\%}$ &
$^{+13.7\%}_{-14.1\%}$ & $^{+13.4\%}_{-4.1\%}$ & $^{+58.1\%}_{-39.8\%}$ &
$10.84$ & $^{+3.9\%}_{-8.7\%}$ & $0.07$ & $^{+53.1\%}_{-39.7\%}$ \\ \hline
$185$ & $0.52$ & $^{+21.1\%}_{-19.4\%}$ & $^{+5.4\%}_{-5.4\%}$ &
$^{+14.0\%}_{-14.3\%}$ & $^{+13.4\%}_{-4.2\%}$ & $^{+58.3\%}_{-40.0\%}$ &
$10.90$ & $^{+3.7\%}_{-8.8\%}$ & $0.06$ & $^{+53.1\%}_{-40.2\%}$ \\ \hline
$190$ & $0.43$ & $^{+21.1\%}_{-19.5\%}$ & $^{+5.4\%}_{-5.4\%}$ &
$^{+14.3\%}_{-14.5\%}$ & $^{+13.5\%}_{-4.2\%}$ & $^{+58.5\%}_{-40.3\%}$ &
$10.95$ & $^{+3.8\%}_{-8.7\%}$ & $0.05$ & $^{+53.3\%}_{-40.4\%}$ \\ \hline
$195$ & $0.36$ & $^{+21.1\%}_{-19.5\%}$ & $^{+5.4\%}_{-5.4\%}$ &
$^{+14.6\%}_{-14.7\%}$ & $^{+13.4\%}_{-4.2\%}$ & $^{+58.8\%}_{-40.5\%}$ &
$11.00$ & $^{+3.8\%}_{-8.8\%}$ & $0.04$ & $^{+53.7\%}_{-40.6\%}$ \\ \hline
$200$ & $0.30$ & $^{+21.1\%}_{-19.6\%}$ & $^{+5.4\%}_{-5.4\%}$ &
$^{+14.9\%}_{-15.0\%}$ & $^{+13.4\%}_{-4.2\%}$ & $^{+59.0\%}_{-40.7\%}$ &
$11.04$ & $^{+3.9\%}_{-8.7\%}$ & $0.03$ & $^{+54.2\%}_{-40.7\%}$ \\ \hline
\end{tabular}
\end{center}
\caption{The production cross sections in the $\protect{gg\to A}$  process at
the Tevatron (in fb) for given $A$ masses (in GeV) at a scale $\mu_F=\mu_R
=\frac12 M_A$ with MSTW PDFs. Shown also are the corresponding uncertainties
from the various sources discussed as well as the total uncertainty. In the
other columns, displayed are the branching ratio BR($A\to \tau^+\tau^-)$ [in \%]
and the product $\sigma(gg\to A)\times$BR$(A\to \tau^+\tau^-)$ together with
their
respective total uncertainties.}
\label{gg-tab}
\vspace*{-1mm}
}
\end{table}

\begin{table}[!h]{\small%
\let\lbr\{\def\{{\char'173}%
\let\rbr\}\def\}{\char'175}%
\renewcommand{\arraystretch}{1.3}
\vspace*{10mm}
\begin{center}
\begin{tabular}{|c||c|cccc|cc||cc|}\hline
$~~M_A~~$ & $\sigma^{\rm NNLO}_{\rm b\bar b \to A}$ & scale &
PDFs & param & total & BR$(\tau\tau)  \hspace*{-5mm}$  &  & $\sigma \times
$BR$ \hspace*{-5mm}$ &
 \\ \hline
$90$ & $26.31$ & $^{+35.1\%}_{-29.9\%}$ &
$^{+18.2\%}_{-17.2\%}$ & $^{+10.5\%}_{-3.3\%}$ & $^{+70.4\%}_{-44.9\%}$ &
$9.60$ & $^{+3.8\%}_{-8.8\%}$ & $2.53$ & $^{+63.4\%}_{-44.0\%}$ \\ \hline
$95$ & $21.04$ & $^{+30.6\%}_{-29.0\%}$ &
$^{+19.0\%}_{-16.6\%}$ & $^{+9.9\%}_{-3.7\%}$ & $^{+64.1\%}_{-44.9\%}$ &
$9.70$ & $^{+3.7\%}_{-8.8\%}$ & $2.04$ & $^{+57.1\%}_{-43.9\%}$ \\ \hline
$100$ & $16.96$ & $^{+26.6\%}_{-27.0\%}$ &
$^{+18.7\%}_{-17.5\%}$ & $^{+10.2\%}_{-3.4\%}$ & $^{+61.5\%}_{-42.9\%}$ &
$9.79$ & $^{+3.8\%}_{-8.8\%}$ & $1.66$ & $^{+54.5\%}_{-42.0\%}$ \\ \hline
$105$ & $13.78$ & $^{+22.6\%}_{-27.2\%}$ &
$^{+19.2\%}_{-17.6\%}$ & $^{+10.7\%}_{-3.4\%}$ & $^{+57.8\%}_{-43.4\%}$ &
$9.88$ & $^{+3.8\%}_{-8.8\%}$ & $1.36$ & $^{+50.8\%}_{-42.4\%}$ \\ \hline
$110$ & $11.22$ & $^{+20.3\%}_{-25.5\%}$ &
$^{+19.4\%}_{-17.9\%}$ & $^{+10.1\%}_{-3.6\%}$ & $^{+54.7\%}_{-42.6\%}$ &
$9.96$ & $^{+3.8\%}_{-8.8\%}$ & $1.12$ & $^{+47.9\%}_{-41.7\%}$ \\ \hline
$115$ & $9.18$ & $^{+17.6\%}_{-24.8\%}$ &
$^{+20.1\%}_{-17.9\%}$ & $^{+10.6\%}_{-3.6\%}$ & $^{+52.7\%}_{-41.6\%}$ &
$10.04$ & $^{+3.8\%}_{-8.8\%}$ & $0.92$ & $^{+45.7\%}_{-40.6\%}$ \\ \hline
$120$ & $7.57$ & $^{+15.1\%}_{-24.1\%}$ &
$^{+21.0\%}_{-17.8\%}$ & $^{+10.9\%}_{-3.0\%}$ & $^{+50.2\%}_{-40.3\%}$ &
$10.12$ & $^{+3.8\%}_{-8.8\%}$ & $0.77$ & $^{+43.1\%}_{-40.0\%}$ \\ \hline
$125$ & $6.29$ & $^{+13.6\%}_{-23.5\%}$ &
$^{+21.5\%}_{-18.1\%}$ & $^{+10.2\%}_{-3.2\%}$ & $^{+47.9\%}_{-40.4\%}$ &
$10.19$ & $^{+3.8\%}_{-8.8\%}$ & $0.64$ & $^{+40.7\%}_{-39.5\%}$ \\ \hline
$130$ & $5.24$ & $^{+12.8\%}_{-22.9\%}$ &
$^{+21.7\%}_{-18.6\%}$ & $^{+10.5\%}_{-3.2\%}$ & $^{+48.0\%}_{-40.4\%}$ &
$10.26$ & $^{+3.8\%}_{-8.7\%}$ & $0.54$ & $^{+41.0\%}_{-39.5\%}$ \\ \hline
$135$ & $4.36$ & $^{+12.4\%}_{-21.9\%}$ &
$^{+21.8\%}_{-19.0\%}$ & $^{+10.5\%}_{-3.3\%}$ & $^{+48.5\%}_{-39.9\%}$ &
$10.33$ & $^{+3.8\%}_{-8.8\%}$ & $0.45$ & $^{+41.6\%}_{-38.9\%}$ \\ \hline
$140$ & $3.66$ & $^{+11.6\%}_{-21.5\%}$ &
$^{+22.5\%}_{-19.1\%}$ & $^{+10.2\%}_{-3.6\%}$ & $^{+46.9\%}_{-40.2\%}$ &
$10.39$ & $^{+3.8\%}_{-8.7\%}$ & $0.38$ & $^{+39.8\%}_{-39.3\%}$ \\ \hline
$145$ & $3.08$ & $^{+11.0\%}_{-21.2\%}$ &
$^{+23.0\%}_{-19.5\%}$ & $^{+10.4\%}_{-3.3\%}$ & $^{+46.7\%}_{-40.3\%}$ &
$10.46$ & $^{+3.7\%}_{-8.8\%}$ & $0.32$ & $^{+39.5\%}_{-39.4\%}$ \\ \hline
$150$ & $2.60$ & $^{+10.8\%}_{-20.3\%}$ &
$^{+24.1\%}_{-19.3\%}$ & $^{+10.4\%}_{-3.4\%}$ & $^{+47.3\%}_{-39.7\%}$ &
$10.52$ & $^{+3.7\%}_{-8.8\%}$ & $0.27$ & $^{+39.9\%}_{-38.9\%}$ \\ \hline
$155$ & $2.20$ & $^{+10.8\%}_{-20.0\%}$ &
$^{+24.3\%}_{-20.1\%}$ & $^{+10.3\%}_{-3.5\%}$ & $^{+47.8\%}_{-39.3\%}$ &
$10.57$ & $^{+3.8\%}_{-8.7\%}$ & $0.23$ & $^{+40.8\%}_{-38.2\%}$ \\ \hline
$160$ & $1.88$ & $^{+10.5\%}_{-20.0\%}$ &
$^{+24.9\%}_{-20.4\%}$ & $^{+10.4\%}_{-3.4\%}$ & $^{+47.8\%}_{-40.0\%}$ &
$10.63$ & $^{+3.8\%}_{-8.7\%}$ & $0.20$ & $^{+40.5\%}_{-39.1\%}$ \\ \hline
$165$ & $1.60$ & $^{+10.6\%}_{-18.7\%}$ &
$^{+25.1\%}_{-20.6\%}$ & $^{+10.4\%}_{-3.3\%}$ & $^{+49.2\%}_{-38.5\%}$ &
$10.68$ & $^{+3.9\%}_{-8.7\%}$ & $0.17$ & $^{+41.8\%}_{-37.4\%}$ \\ \hline
$170$ & $1.37$ & $^{+10.6\%}_{-18.4\%}$ &
$^{+25.3\%}_{-21.1\%}$ & $^{+10.5\%}_{-3.1\%}$ & $^{+49.8\%}_{-38.7\%}$ &
$10.74$ & $^{+3.8\%}_{-8.8\%}$ & $0.15$ & $^{+42.4\%}_{-37.8\%}$ \\ \hline
$175$ & $1.17$ & $^{+10.5\%}_{-18.2\%}$ &
$^{+26.0\%}_{-21.5\%}$ & $^{+10.4\%}_{-3.4\%}$ & $^{+50.5\%}_{-38.8\%}$ &
$10.79$ & $^{+3.8\%}_{-8.7\%}$ & $0.13$ & $^{+43.0\%}_{-37.6\%}$ \\ \hline
$180$ & $1.01$ & $^{+10.3\%}_{-18.1\%}$ &
$^{+26.7\%}_{-21.6\%}$ & $^{+10.4\%}_{-3.4\%}$ & $^{+50.7\%}_{-38.6\%}$ &
$10.84$ & $^{+3.9\%}_{-8.7\%}$ & $0.11$ & $^{+43.2\%}_{-37.4\%}$ \\ \hline
$185$ & $0.87$ & $^{+10.4\%}_{-17.6\%}$ &
$^{+27.8\%}_{-21.5\%}$ & $^{+10.3\%}_{-3.2\%}$ & $^{+51.4\%}_{-38.6\%}$ &
$10.90$ & $^{+3.7\%}_{-8.8\%}$ & $0.09$ & $^{+43.8\%}_{-37.6\%}$ \\ \hline
$190$ & $0.75$ & $^{+10.3\%}_{-17.3\%}$ &
$^{+27.2\%}_{-22.7\%}$ & $^{+10.4\%}_{-3.3\%}$ & $^{+52.1\%}_{-39.2\%}$ &
$10.95$ & $^{+3.8\%}_{-8.7\%}$ & $0.08$ & $^{+44.4\%}_{-38.2\%}$ \\ \hline
$195$ & $0.65$ & $^{+10.5\%}_{-16.2\%}$ &
$^{+28.5\%}_{-22.6\%}$ & $^{+10.1\%}_{-3.4\%}$ & $^{+52.3\%}_{-38.9\%}$ &
$11.00$ & $^{+3.8\%}_{-8.8\%}$ & $0.07$ & $^{+44.7\%}_{-37.7\%}$ \\ \hline
$200$ & $0.56$ & $^{+10.4\%}_{-16.3\%}$ &
$^{+28.5\%}_{-23.6\%}$ & $^{+10.6\%}_{-3.2\%}$ & $^{+53.2\%}_{-38.9\%}$ &
$11.04$ & $^{+3.9\%}_{-8.7\%}$ & $0.06$ & $^{+45.8\%}_{-37.9\%}$ \\ \hline
\end{tabular}
\end{center}
\caption{The production cross sections in the $\protect{b\bar b\to A}$ 
process at
the Tevatron (in fb) for given $A$ masses (in GeV) at a scale $\mu_F=\mu_R
=\frac14 M_A$ with MSTW PDFs. Shown also are the corresponding uncertainties
from the various sources discussed as well as the total uncertainty. In the
other columns, displayed are the branching ratio BR($A\to \tau^+\tau^-)$ 
[in \%] and the product $\sigma(b\bar b\to A)\times$BR$(A\to \tau^+\tau^-)$ 
together with their respective total uncertainties.}
\label{bb-tab}
\vspace*{-1mm}
}
\end{table}

\end{document}